%% file: main.tex
\documentclass[final,journal,a4paper,twocolumn]{IEEEtran}

\usepackage{hyperref}   
\hypersetup{colorlinks=false,linkbordercolor=.75 .9 .95,citebordercolor=.75 .9 .95,urlbordercolor=.5 .7 .1}
\usepackage{cite}
\usepackage{amsmath,amssymb,amsfonts}
\usepackage{algorithmic}
\usepackage{graphicx}
\usepackage{textcomp}
\usepackage{xcolor}
\usepackage{enumitem}
\usepackage{subfigure} 
\usepackage{multirow}
\RequirePackage{bm}
\RequirePackage{mathrsfs}
\usepackage{booktabs} 
\usepackage{threeparttable}
\usepackage{diagbox} 
\graphicspath{{figures/}}
\usepackage{CJKutf8}

\makeatletter
\renewcommand{\maketag@@@}[1]{\hbox{\m@th\normalsize\normalfont#1}}
\makeatother

\usepackage{float}
\usepackage{algorithm}
\floatstyle{plain} 
\newfloat{myairfloat}{htbp}{myair}
\newenvironment{myAlgo}[2][htb]%
{	\begin{myairfloat}[#1]
		\centering
		\begin{minipage}{#2}
			\begin{algorithm}[H]}
			{\end{algorithm}
		\end{minipage}
	\vspace{-5mm}
	\end{myairfloat}
}
\allowdisplaybreaks[4]
\newlength{\Oldarrayrulewidth}

\def\BibTeX{{\rm B\kern-.05em{\sc i\kern-.025em b}\kern-.08em
    T\kern-.1667em\lower.7ex\hbox{E}\kern-.125emX}}

\newtheorem{theorem}{\bf Proposition}

\newtheorem{remark}{Remark}

\begin{document}
\bstctlcite{myrefs:BSTcontrol}
% \title{RSSI-Based Localization Utilizing Antenna Radiation Pattern And Biased CRLB Analysis }
% \title{Low-Cost RSSI Positioning Using Single-Antenna Radiation Pattern Variations}
\title{Cost-Effective Single-Antenna RSSI Positioning Through Dynamic Radiation Pattern Analysis}
\author{Zhisheng Rong\textsuperscript{*}, Wenzhi Liu\textsuperscript{*}, Xiayue Liu, Zhixiang Xu,\\ Yufei Jiang,~\IEEEmembership{Member,~IEEE}  Xu Zhu,~\IEEEmembership{Senior Member,~IEEE,} 
   \thanks{*These authors contributed equally to this work.}
}
\maketitle
\begin{CJK}{UTF8}{gbsn}
        \input{body/C1}
        \input{body/C2}

        \input{body/C3}

        \input{body/C4}
        \input{body/C5}

        \input{body/appendix}
\end{CJK}

% \bibliographystyle{IEEEtran}
% \bibliography{IEEEabrv,myrefs}
\input{main.bbl}

% \bibliographystyle{IEEEtran}
% \bibliography{myrefs}

\end{document}

%% file: body/C1.tex
\begin{abstract}
Mobile robot positioning systems traditionally rely on multiple antennas, increasing hardware complexity and cost. Thus, we propose a single-antenna positioning approach, where  radiation pattern characteristics are leveraged based on maximum likelihood estimation (MLE) through received signal strength indication (RSSI) measurements. The proposed positioning approach is cost-effective, requiring only a single antenna. By rotating the antenna, we exploit asymmetric radiation patterns to achieve pattern diversity, which allows the single-antenna positioning system to approach the performance of multi-point positioning systems. 
We further propose a geometric-based curve intersection detection (CID) algorithm that outperforms conventional equation-solving methods. We theoretically derive the biased Cramér-Rao lower bound (CRLB) of the proposed positioning approach, dependent on signal-to-noise ratio (SNR), directivity of antenna radiation pattern and the number of rotation measurements. Simulation results show that all methods improve with increasing SNR, number of rotation and directivity of antenna radiation pattern, and the MLE achieves performance close to the derived biased CRLB.

\end{abstract}

\vspace{-3mm}
\section{Introduction}
% designed for indoor environments 
Mobile robots play a crucial role in various fields, including industrial automation, smart logistics, emergency response, environmental monitoring, and healthcare\cite{ullah2024mobile}. 
Accurate positioning are critical for the effective operation of mobile robots. While Global Positioning System (GPS) and similar satellite-based technologies have transformed outdoor navigation, their positioning accuracy remains insufficient for robot location \cite{farahsari2022survey}. 
Current research in robotics has primarily relied on simultaneous localization and mapping (SLAM) , which suffer from computational complexity and trajectory constraints \cite{ullah2024mobile}. Consequently, robot wireless positioning techniques have emerged as a promising research frontier to enhance mobile robot autonomy \cite{farahsari2022survey}.

Among the various positioning techniques, technologies like Time of Flight (ToF), Angle of Arrival (AoA), Phase of Arrival (PoA), Received Signal Strength Indication (RSSI), and Channel State Information (CSI) have been extensively studied \cite{mu2021intelligent}.  While each approach has its own limitations.  ToF requires precise timing, AoA struggles with obstacles, PoA is noise-sensitive, and CSI demands extensive data processing. In contrast, RSSI stands out as a more practical option due to its simpler hardware requirements and easier integration with existing wireless systems, even though its positioning accuracy is relatively lower \cite{singh2022comparison}. 

Many researchers have focused on improving the accuracy and applicability of RSSI for robot position. Approaches such as filtering algorithms and machine learning have been used to enhance positioning accuracy and reduce errors\cite{hoang2019recurrent}. However, these methods often increase computational complexity, require large datasets, and introduce delays that impact real-time performance. As a result, RSSI is still primarily used as an auxiliary tool in robot position, such as distance estimation or database creation \cite{yucel2020wi}. 

Newer approaches aim to expand RSSI's potential by addressing signal fluctuations at the transmitter side. For example, \cite{li2018indoor} proposed an positioning method that updates stored data with real-time RSSI values, changing the parameters to reduce the result error according to the fluctuation of the transmitter.
Other studies challenge the assumption of omnidirectional antenna radiation patterns, considering the impact of radiation pattern on signal strength. Research has shown that antenna polarization and irregular radiation patterns affect positioning accuracy \cite{maeng2023impact}, and incorporating radiation pattern into positioning algorithms\cite{mwila2014use} or using positioning techniques based on radiation pattern \cite{wang2019arpap} has led to improved accuracy. However, many of these studies analyze radiation pattern as a source of error, failing to fully exploit the additional information provided by them. 

Building upon prior research, we develop a novel single-antenna positioning system, leveraging the inherent asymmetry of antenna radiation patterns. The positioning capability is realized by dynamically controlling antenna radiation pattern variations, implemented either via mechanical antenna rotation or reconfigurable antenna. Compared to conventional multi-antenna positioning systems, the proposed single-antenna approach offers notable advantages, including reduced hardware complexity and lowered design expenses, while preserving positioning accuracy.
The main contributions of this paper are summarized as follows:
\begin{itemize}
    \item We propose a localization scheme leveraging pattern diversity through antenna rotation, where asymmetric radiation patterns enable single-antenna systems to approach multi-antenna performance. Based on this approach, we derive a maximum likelihood estimation (MLE) positioning method that approaches the derived biased Cramér-Rao lower bound (CRLB), showing superior distance estimation but inferior angle estimation compared to multi-point systems.
    \item We establish a mathematical framework characterizing estimation accuracy bounds and analyze system performance. Our analysis reveals that positioning accuracy improves with increased SNR, more antenna rotations, and steeper radiation pattern variations, providing essential design guidelines for practical implementation.
    \item We develop a generalized geometric algorithm called curve intersection detection (CID) for robust localization under both antenna rotation and reconfigurable pattern variations scenarios. This method achieves estimation accuracy closer to the CRLB compared to equation-solving approaches.
\end{itemize}

%% file: body/C2.tex
% !TeX spellcheck = en_US
% !TEX root = ../conference_main.tex

\section{System Model}
\begin{figure}
        \centering
        \includegraphics[width=1\linewidth]{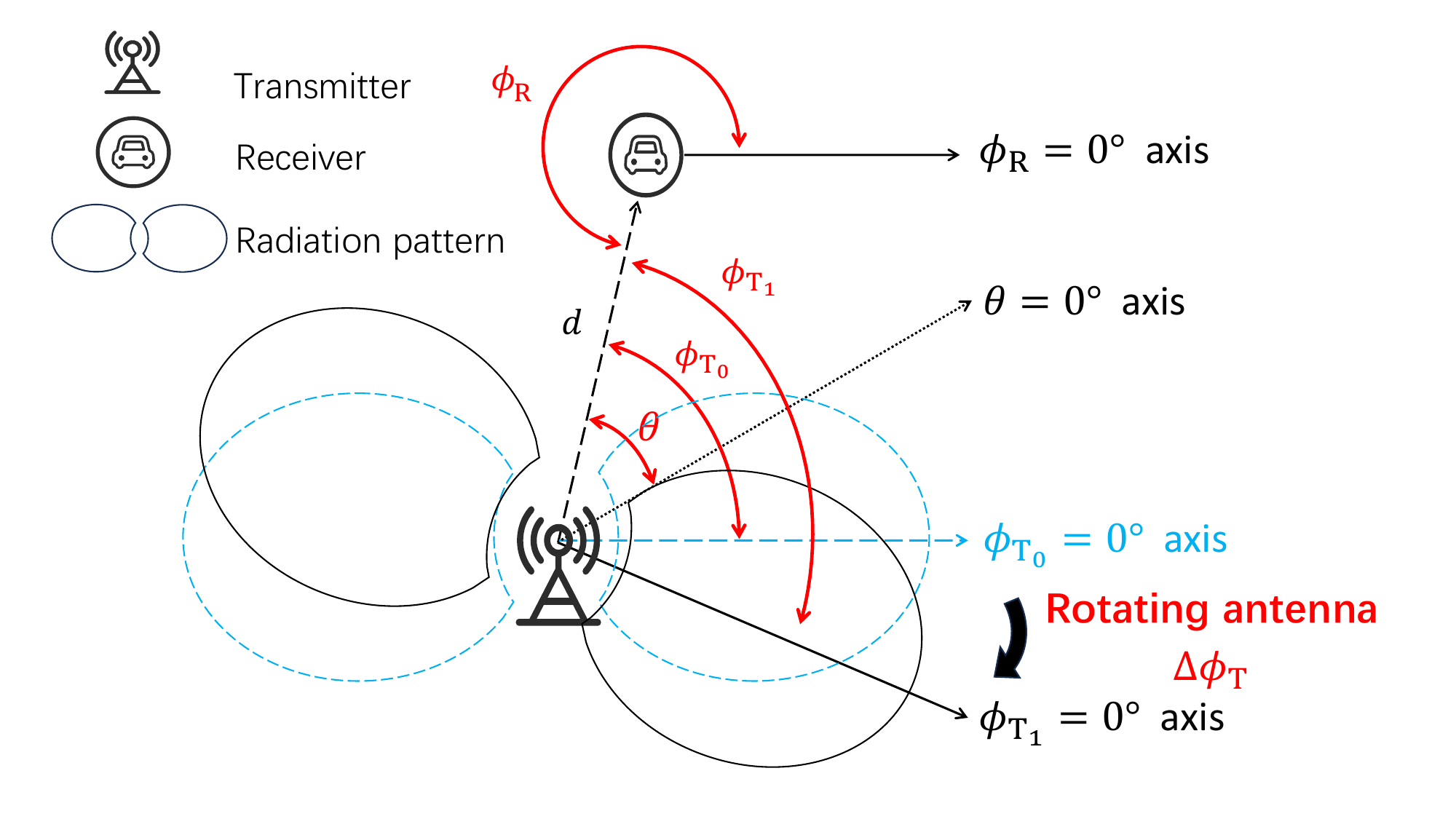}
        \caption{ System model for single-antenna positioning using pattern rotation, where $d$ denotes transmitter-receiver distance , and $\theta$ represents the angle relative to the $\theta = 0^\circ$ axis. 
        $\bigtriangleup\phi_{\textrm{T}}$ denotes the angle the axis changes from  $\phi_{\textrm{T}_0} = 0^\circ$ to $\phi_{\textrm{T}_1} = 0^\circ$, indicating the antenna rotation. And $\phi_{\textrm{T}_i}$ represents the angle between the transmitter and receiver in the $i$-th transmitting antenna coordinate system relative to $\phi_{\textrm{T}_i} = 0^\circ$ axis, $\phi_{\textrm{R}}$ represents the angle  in the receiving antenna coordinate system relative to its $\phi_{\textrm{R}} = 0^\circ$ axis.
        }
        \label{fig:sys-model}
\end{figure}

Consider a robotic swarm communication, where each robotic equipped with a single antenna. This configuration is particularly relevant to the system consisting of a single transmitter and multiple receivers\cite{ullah2024mobile,farahsari2022survey}, as shown in  Fig.\ref{fig:sys-model}. 

In this paper, we assume the capability of rotating antenna radiation patterns through either mechanical rotation or electronic steering \cite{9743796}.   
As shown in  Fig.\ref{fig:sys-model}, $\phi_{\textrm{T}_i}$ denotes the transmission angle measured from the transmitter's reference axis after the antenna has undergone $i$ rotations, and $\phi_{\textrm{T}_0}$ denotes the the initial transmission angle before any rotation relative to $\phi_{\textrm{T}_0} = 0^{\circ}$ axis. $\phi_{\textrm{R}}$ denotes the reception angle measured from the receiver's reference axis, and θ represents the relative angle between transmitter and receiver. $d$ denotes the distance between transmitter and receiver.
The known angle of the transmitting antenna radiation pattern rotations is defined as $\Delta\phi_{\textrm{T}}$.
For simplicity of expression, we align $\phi_{\textrm{T}_0} = 0^{\circ}$ axis with the $\theta = 0^{\circ}$ axis without loss of generality, thus
\begin{equation} \label{111}
\small
    \phi_{\textrm{T}_0} = \theta,\quad \phi_{\textrm{T}} = \theta + \bigtriangleup\phi_{\textrm{T}}.
\end{equation}

According to the Friis transmission equation \cite{shaw2013radiometry} and considering the shadowing fading effects, the RSSI  between transmitter and receiver positioned at $(d,\theta)$ relative to transmitter can be expressed as
\begin{equation}
\small
\mathcal{R}(\phi_{\textrm{T}},\phi_{\textrm{R}})|_{(d,\theta)} = 10 \log_{10} \left(\frac{}{} \frac{P_{\textrm{T}} \lambda^2 G_{\textrm{T}}(\phi_{\textrm{T}})  G_{\textrm{R}}{(\phi_{\textrm{R}})} }{(4\pi)^2d^n}\right) + 30 + X_\sigma  \label{eq:R},
\end{equation}
where $P_{\textrm{T}}$ denotes the transmission power, $\lambda$ denotes the signal wavelength. $G_{\textrm{T}}(\phi_{\textrm{T}})$ and $G_{\textrm{R}}(\phi_{\textrm{R}})$  denote the antenna gains at the transmitter and receiver, respectively. 
The path loss exponent $n$ characterizes the propagation environment,
and is usually a number from 2 to 6 \cite{rappaport2024wireless}. 
$X_\sigma$ represents the shadowing effects, modeled as a zero-mean Gaussian random variable with variance $\sigma^2$, \textit{i.e.},  $X_\sigma \sim \mathcal{N}(0,\sigma^2)$.

%% file: body/C3.tex
\section{Stationary Positioning}
To achieve RSSI-based positioning with a single antenna, we conducting theoretical analysis, leveraging the radiation pattern rotation capabilities of reconfigurable antennas. Subsection \ref{book:ABC} analyzes the feasibility of this positioning approach with omnidirectional receiving antennas, featuring a  MLE method along with comprehensive performance evaluation, and derives the biased CRLB .  Subsection \ref{subsection:BBC} extends this approach by applying two-position measurement strategy to address positioning challenges with directional receiving antennas.
\subsection{Under Known Receiving Antenna Radiation Pattern} \label{book:ABC}
For a fixed receiving antenna at position $(d_0,\theta_0)$ with constant radiation pattern $G_{\textrm{R}}$, $d_0$ can be derived from (\ref{eq:R})
% a function of the antenna angles $\phi_{\textrm{T}}$  from ($\ref{eq:R}$):
\begin{equation} \label{target curve}
 d_0\!=\!\sqrt[n]{A^2 G_{\textrm{T}}(\phi_{\textrm{T}}) G_{\textrm{R}}}\cdot{10^{-\frac{\mathcal{R}(\phi_{\textrm{T}},\phi_{\textrm{R}})|_{(d_0,\theta_0)}-30-X_{\sigma}}{10 n}}}.
\end{equation} 
where $A = \sqrt{\frac{P_{\textrm{T}} \cdot \lambda}{(4\pi)^2}}$.
This formulation yields a system of coupled equations in $d_0$ and $\phi_\textrm{T}$ .
 
 To solve this coupled system, we rotate the transmitting antenna through different angles [$\phi_{\textrm{T}_0},\phi_{\textrm{T}_1},...,\phi_{\textrm{T}_N}$] which provides a sequence of signal strength measurements [$R(\phi_{\textrm{T}_0})$,$R (\phi_{\textrm{T}_1})$,...,$R (\phi_{\textrm{T}_N})$] given by:
\begin{align}
\small
        R(\phi_{\textrm{T}_i}) = 10\cdot\log_{10} (A^2 \frac{{{G_{\textrm{T}}(\phi_{\textrm{T}_i}}}) G_{\textrm{R}}}{{d_0}^n}) + 30 + X_{{\sigma}_i} \label{eq:Ri},
\end{align}
where $X_{{\sigma}_i} \sim \mathcal{N}(0,\sigma^2)$. Since $M_i(d,\theta) = 30 + 10n \cdot \log_{10}(\frac{\sqrt[n]{A^2G_{\textrm{T}_i}(\theta+\bigtriangleup\phi_{\textrm{T}})G_{\textrm{R}}}}{d_0})$ is a constant, $R(\phi_{\textrm{T}_i})$ is a Gaussian variable with mean $ M_i(d,\theta)$ and variance $\sigma^2$,\textit{i.e.}
\begin{align}
\small
    R(\phi_{\textrm{T}_i}) \sim \mathcal{N}\left(M_i(d,\theta),\sigma^2 \right).
\end{align}

Since the angular differences between antenna pattern rotations [$\bigtriangleup\phi_{\textrm{T}_1}$,$\bigtriangleup\phi_{\textrm{T}_2}$,...,$\bigtriangleup\phi_{\textrm{T}_N}$] are known, we can propose $N$ equations from (\ref{target curve}):
\begin{equation} \label{eq:solve}
	d_0 =\sqrt[n]{{A^2G_{\textrm{T}}(\theta_0+\bigtriangleup\phi_{\textrm{T}_i})\cdot G_{\textrm{R}}}}\cdot10^{-\frac{R(\phi_{\textrm{T}_i})-30-X_{{\sigma}_i}}{10n}}
\end{equation}
where $\bigtriangleup\phi_{\textrm{T}_i} = \phi_{\textrm{T}_i}-\phi_{\textrm{T}_0}$, and $i$ from 1 to $N$. 

\begin{myAlgo}[t]{0.45\textwidth}  
	% \caption{A localization algorithm with known radiation patterns of the receiving and transmitting antennas}  
        \caption{Curve Intersection Detection (CID)}
	\label{algo:angle_locate}
	\begin{algorithmic}[1] %这个1 表示每一行都显示数字
		\REQUIRE ~~  %算法的输入参数：Input
		$n_{\text{t}},P_{\text{t}},G_{\text{t}},G_{\text{r}},d_0,\theta_0$;
		\ENSURE ~~\\ 
		\FOR{$i=1$ \TO $N$}
            \STATE Given $R(\phi_{\textrm{T} i})$ is obtained by (\ref{eq:Ri});
		\STATE Given target curve $d(\theta)|_{\phi_{\textrm{T}} = {\phi_{\textrm{T}_i}}}$  is obtained by (\ref{eq:solve});
		\ENDFOR	
		\STATE Given ${(d,\theta)_{ij}}$ is obtained from the intersection points of the curves $d(\theta)|_{\phi_{\textrm{T}} = \phi_{\textrm{T}_i},\phi_{\textrm{Tj}}}$;
		\STATE Given $(\hat{d},\hat{\theta})$ is obtained from the average value of ${(d,\theta)_{ij}}$;
		\RETURN $(\hat{d},\hat{\theta})$; %算法的返回值
	\end{algorithmic}
\end{myAlgo}  
$d_0$ and $\theta_0$ can be determined by solving the system of equations (\ref{eq:solve}), though the nonlinear solution process incurs significant computational cost. 
Then a Curve Intersection Detection Algorithm, \textbf{Algorithm \ref{algo:angle_locate}}, is proposed to determine $d$ and $\theta$ efficiently. This algorithm identifies critical intersection points of distinct function curves, where each curve comes from the equation in (\ref{eq:solve}) and has the form shown in (\ref{target curve}) with respect to $\phi_\textrm{T}$. Therefore, these intersection points represent the solutions to the corresponding equations. This method avoids the complex nonlinear system solving process, and is applicable to both antenna rotation and reconfigurable antenna pattern variations since (\ref{target curve}) imposes no constraints on antenna radiation patterns.

For the case of antenna rotation, the radiation pattern remains consistent across all angular positions.
We propose an estimator based on the MLE methodology to improve both computational efficiency and estimation accuracy,  
% 【什么超级长难句，改一下】

Given the observation $R(\phi_{\textrm{T}_i})$ to estimate its probability density function, the log-likelihood function $L(d,\theta⋅)$ is defined as
% \begin{equation}
% \small
%     L(d,\theta⋅) = \prod_{i=1}^{N} \frac{1}{\sigma\sqrt{2\pi}}e^{-\frac{[R(\phi_{\textrm{T}_i})-M_i(d,\theta)]^2}{2\sigma^2}} .
% \end{equation}
% The likelihood function is as:
\begin{align}
\small
    \ln(L(d,\theta)) = -\frac{N}{2}\ln(2&\pi\sigma^2) - \frac{1}{2\sigma^2}\sum_{i=1}^{N}[R(\phi_{\textrm{T}_i}) - M_i(d,\theta)]^2.
\end{align} 

The optimal parameters $(\hat{d},\hat{\theta})$ can be determined by maximizing the log-likelihood function $\ln(L(d,\theta))$, \textit{i.e.}, 
\begin{align}
    (\hat{d},\hat{\theta}) &= \arg\max_{(d,\theta)} \ln(L(d,\theta)) .
\end{align}
To obtain the optimal values of $(\hat{d},\hat{\theta})$ that maximize $\ln(L(d,\theta))$, we solve the following system of equations
\begin{align}
\tiny  
    \frac{\partial \ln L(d,\theta)}{\partial d}\!=\!\sum_{i=1}^{N} \left([R(\phi_{\textrm{T}_i}) - M_i(\hat{d}, \hat{\theta})]\frac{\partial M_i(d, \theta)}{\partial d} \right) = 0 \label{eq:over_d}\\ 
    \frac{\partial \ln L(d, \theta)}{\partial \theta}\!=\!\sum_{i=1}^{N} \left([R(\phi_{\textrm{T}_i}) - M_i(\hat{d}, \hat{\theta})] \frac{\partial M_i(d, \theta)}{\partial \theta} \right)= 0 \label{eq:over_theta},
\end{align}
$\hat{d}$ and $\hat{\theta}$ can be solved from the nonlinear equations (\ref{eq:over_d})(\ref{eq:over_theta}).

\begin{theorem} \label{theory:hat_d}
The distance estimator $\hat{d}$ \eqref{eq:E(d)} is biased. The bias level increases with noise intensity $\sigma^2$ and angular estimation $\hat{\theta}$, while decreasing with the number of measurements $N$.
	\vspace{-2mm}
		\begin{align} 
            \small
			 \hat{d} = d_0 \cdot 10^{\frac{X_\sigma ^{'}}{10n \cdot N} } \cdot \left(\prod_{i=1}^{N}\Gamma_i(\hat{\theta}) \right)^{\frac{1}{n\cdot N}}\label{eq:E(d)},
		\end{align}
	\vspace{-3mm}
\end{theorem}
where $X_\sigma ^{'} =- \sum_{i=1}^{N} X_{{\sigma}_i}$, $\Gamma_i(\hat{\theta}) = {{\frac{{{G_{\textrm{T}}(\hat{\theta}+\bigtriangleup\phi_{\textrm{T}_i})}}}{{{G_{\textrm{T}}(\theta_0+\bigtriangleup\phi_{\textrm{T}_i})}}}}}$. Notably, the distance estimate maintains its bias even when the angular estimator $\hat{\theta}$ is unbiased. This persistent bias stems from the inherent properties of the log-normal distribution, specifically that its mean value deviates from unity.
\begin{IEEEproof}
      See Appendix \ref{book:hat_d}  
\end{IEEEproof}

\begin{theorem}  \label{theory:hat_theta}
The angular estimator $\hat{\theta}$ solved by (\ref{eq:11}) is independent of the distance parameter $d$,  : 
{\small
     \begin{align} 
     10n \log_{10}\left(\prod_{i=1}^{N}  \Gamma_i(\hat{\theta})^{\frac{\mathcal{H}-N\cdot K_i}{n\cdot N}}\right)\!+\!\sum_{i=1}^{N} (K_i-\frac{\mathcal{H}}{N} )X_{{\sigma}_i}\!=\!0 \label{eq:11}.
     \end{align}
}
where $K_i= \frac{10}{\ln10}\cdot \frac{G_{\textrm{T}}^{'}(\hat{\theta}+\bigtriangleup\phi_{\textrm{T}_i})}{G_{\textrm{T}}(\hat{\theta}+\bigtriangleup\phi_{\textrm{T}_i})}$, $\mathcal{H} = \sum_{i=1}^{N}K_i$. 
\end{theorem}
\begin{IEEEproof}
See Appendix \ref{book:hat_theta}
\begin{remark}
Furthermore, the accuracy of angular estimation improves with increasing  variance of the sequence  $\{\frac{G_{\textrm{T}}^{'}(\theta+\bigtriangleup\phi_{\textrm{T}i})}{G_{\textrm{T}}(\theta+\bigtriangleup\phi_{\textrm{T}_i})}\}_{i=1}^{N}$, denoted as $\mathcal{G}$.  A larger variance of $\mathcal{G}$ indicates increased directivity in the antenna radiation pattern.
\end{remark}	
\end{IEEEproof}

\begin{theorem} \label{theory:CRLB}
The CRLB for biased distance estimation $\hat{d}$ and angular estimation $\hat{\theta}$ is derived as:
\begin{align} \label{geq:Cov}
\small
\text{CRLB}({[\hat d,\hat \theta]}) = 
\begin{bmatrix}
\beta_d^2 & \beta_d \beta_\theta \\
\beta_d \beta_\theta & \beta_\theta^2
\end{bmatrix}
+
\begin{bmatrix}
\frac{\partial \beta_d}{\partial d} & \frac{\partial\beta_d}{\partial \theta} \\
\frac{\partial \beta_\theta}{\partial d} & \frac{\partial \beta_\theta}{\partial \theta}
\end{bmatrix}
\mathbf{J}_D^{-1}
\begin{bmatrix}
\frac{\partial \beta_d}{\partial d} & \frac{\partial\beta_d}{\partial \theta} \\
\frac{\partial \beta_\theta}{\partial d} & \frac{\partial \beta_\theta}{\partial \theta}
\end{bmatrix},
\end{align}
where：
\begin{align}
\small
\mathbf{J}_D^{-1} = 
\begin{bmatrix}
\mathbf{J}_D^{-1}{}_{1,1} & \mathbf{J}_D^{-1}{}_{1,2} \\
\mathbf{J}_D^{-1}{}_{2,1} & \mathbf{J}_D^{-1}{}_{2,2} 
\end{bmatrix} \notag,
\end{align}
\begin{align}
\small
& \left[\mathbf{J}_D^{-1}\right]_{1,1} =\frac{\frac{\sigma^2 d^2 \ln^2 10}{100n^2} \cdot \sum_{i=1}^{N} K_i^2}{N \cdot \sum_{i=1}^{N} K_i^2 -
\left( \sum_{i=1}^{N} K_i\right)^2} \label{eq:CRLB_d} ,\\
&\left[\mathbf{J}_D^{-1}\right]_{2,2}=\frac{ N \sigma^2 }{N\cdot \sum_{i=1}^{N} K_i^2 -
\left( \sum_{i=1}^{N} K_i \right)^2} ,\label{eq:CRLB_theta}\\
&\left[\mathbf{J}_D^{-1}\right]_{1,2} = \left[\mathbf{J}_D^{-1}\right]_{2,1}=
\frac{\frac{\sigma^2 d \ln 10}{10n} \cdot \sum_{i=1}^{N} K_i^2}{N \cdot \sum_{i=1}^{N} K_i^2 -
\left( \sum_{i=1}^{N} K_i\right)^2} \label{eq:J21}.
\end{align}

\end{theorem}
\begin{IEEEproof}
See Appendix \ref{book:CRLB}
\end{IEEEproof}
\begin{remark}
Based on (\ref{eq:CRLB_d}) and (\ref{eq:CRLB_theta}), we deduce that
$\textrm{CRLB}(\hat{d})$ increases  with noise intensity $\sigma^2$ and distance parameter $d$, while decreasing with the number of measurements $N$.
Conversely, $\textrm{CRLB}(\hat{\theta})$ increases with $\sigma^2$ and decreases with $N$, but remains unaffected by $d$.
And for high signal-to-noise ratio (SNR) scenarios, where the estimates of $\hat{d}$ and $\hat{\theta}$ exhibit negligible bias, we derive the lower bound for the mean squared errors which are expressed as MSE($\hat{d}$) and MSE($\hat{\theta}$)
% 在d和、theta估计无偏的情况下，或偏差很小，这常常发生在高信噪比情况下，可以直接得到
\begin{equation}
    \textrm{MSE}(\hat{d}) \geq \left[\mathbf{J}_D^{-1}\right]_{1,1},
    \textrm{MSE}(\hat{\theta}) \geq \left[\mathbf{J}_D^{-1}\right]_{2,2} \label{eq:CRLB}.
\end{equation}
\end{remark}

\subsection{Under Unknown Receiving Antenna Radiation Pattern} \label{subsection:BBC}
While Subsection \ref{book:ABC} addresses position estimation with known antenna radiation patterns, practical scenarios often encounter unknown receiving antenna patterns $G_{\textrm{R}}(\phi_{\textrm{R}})$, necessitating an enhanced positioning strategy.

We eliminate the dependence on $G_{\textrm{R}}(\phi_{\textrm{R}})$ by exploiting the differential RSSI measurements between two distinct angles. From (\ref{eq:R}), this relationship is expressed as:
\begin{equation}
\small
\mathcal{R}(\phi_{\textrm{T}_1}, \phi_{\textrm{R}}) - \mathcal{R}(\phi_{\textrm{T}_2}, \phi_{\textrm{R}}) = 10 \log_{10}\left( \frac{G_{\textrm{T}}(\phi_{\textrm{T}_1})}{G_{\textrm{T}}(\phi_{\textrm{T}_2})} \right)+X_{\sigma_1}-X_{\sigma_2}.
\end{equation}
This formulation reveals that the RSSI differential corresponds to the transmitting antenna gain ratio. We characterize this relationship by varying $\phi_{\textrm{T}}$,
\begin{align} \label{eq:ratio}
	r(\phi_{\textrm{T}}) &= 10^{\frac{1}{10} (\mathcal{R}(\theta_0 + \bigtriangleup\phi_{\textrm{T}}, \phi_{\textrm{R}}) - \mathcal{R}(\theta_0, \phi_{\textrm{R}}))}\\
    &= \frac{G_{\textrm{T}}(\theta_0+\bigtriangleup \phi_{\textrm{T}})}{G_{\textrm{T}}(\theta_0)} \cdot 10^{\frac{X_{\sigma_2}-X_{\sigma_1}}{10}}.
\end{align} 
The initial angle $\theta_0$ is estimated by correlating the measured curve with the known normalized transmitting radiation pattern $\frac{G_{\textrm{T}}(\theta_0+\phi_{\textrm{T}})}{G_{\textrm{T}}(\theta_0)}$. 

 We employ similarity analysis as detailed in \textbf{Algorithm \ref{algo: similarity analysis}} to determine the object's direction $\theta_0$
to mitigate the effects of noise and potential reference gain misalignment.
And a two-position measurement strategy is used in order to overcome the unknown receiving antenna gain problem. 
After removing the transmitting antenna by a distance $m$  and obtaining a second angular measurement $\theta_1$ of receiving antenna, geometric relationships yield:
\begin{equation}
d_0 = \frac{\sin(\theta_1)\cdot m}{\sin(\theta_1-\theta_0)} ,
d_1 = \frac{\sin(\theta_0)\cdot m}{\sin(\theta_1-\theta_0)} .
\end{equation}
\begin{myAlgo}[t]{0.45\textwidth}  %h for here
    \caption{Angular Similarity-Based Localization}
    \label{algo: similarity analysis}
    \begin{algorithmic}[1] %这个1 表示每一行都显示数字
	\REQUIRE ~~  %算法的输入参数：Input
	$n_{\text{t}},P_{\text{t}},G_{\text{t}},G_{\text{r}},d_0,\theta_0$;
	\ENSURE ~~\\  
	\FOR{$i=1$ \TO $N$}
	\STATE Given $\mathcal{R}(\phi_{\textrm{T}},\phi_{\textrm{R}})|_{(d_0,\theta_0)}$ is obtained by (\ref{eq:R});
	\STATE Given $r(\phi_{\textrm{T}})$  is obtained by (\ref{eq:ratio});
	\ENDFOR	
	\STATE Given $\hat{\theta}$ is obtained from the similarity estimation between the curves $r(\phi_{\textrm{T}})$ and  $\frac{G_{\textrm{T}}(\theta+\bigtriangleup\phi_{\textrm{T}})}{G_{\textrm{T}}(\theta)}$;
	\STATE Given $\hat{d}$ is obtained by (\ref{target curve})
\RETURN $(\hat{d},\hat{\theta})$; %算法的返回值
\end{algorithmic}
\end{myAlgo}

%% file: body/C4.tex
\section{Simulation Results} \label{book:res}

\subsection{Simulation Setup}
In this section, we conduct numerical simulations to verify the theoretical analysis and the proposed algorithms. We model a half-dipole antenna as 
the transmitting antenna and operate in a 2D plane to satisfy directional pattern requirements. The path loss exponent in the Friis equation is $n = 4$ to emulate indoor environments. The transmission parameters are configured with $P_{\rm T} = 100$ mW and $\lambda = 125$ mm. 
An omnidirectional antenna with gain $G_{\rm R} = 0$ dB serves as the receiving antenna, deployed within the range of 0.5 - 5 m and $-70^{\circ}$ to $70^{\circ}$.
The scenario settings in Figs. \ref{fig:CRLB_d}, \ref{fig:CRLB_theta} are as follows:
the transmitter rotation count is $N = 8$  with a step angle of $\bigtriangleup \phi_{\rm T} = 4^\circ$, while the SNR varies from 0 to 20 dB. 
The scenario settings in Figs. \ref{fig:N_d}, \ref{fig:N_theta} are as follows:
SNR = 10 dB, $\bigtriangleup \phi_{\rm T} = 3^\circ$, while $N$ varies from  = 5 to 14.
And in Fig. \ref{fig:deltaphi_theta}:
SNR = 10 dB, $N = 8$, while  $\bigtriangleup \phi_{\rm T}$ varies from $1^{\circ}$ to $10^{\circ}$ to analysis the effects of  $\mathcal{G}$ variance.

\subsection{Estimation Accuracy Comparison}
We evaluate the positioning performance  by analyzing the mean square error (MSE) of the estimates under various operational conditions. 

\begin{figure}[t]
\subfigure[Distance estimation $d$]{
	\begin{minipage}[t]{0.50\linewidth}
	{\includegraphics[width=1\textwidth]{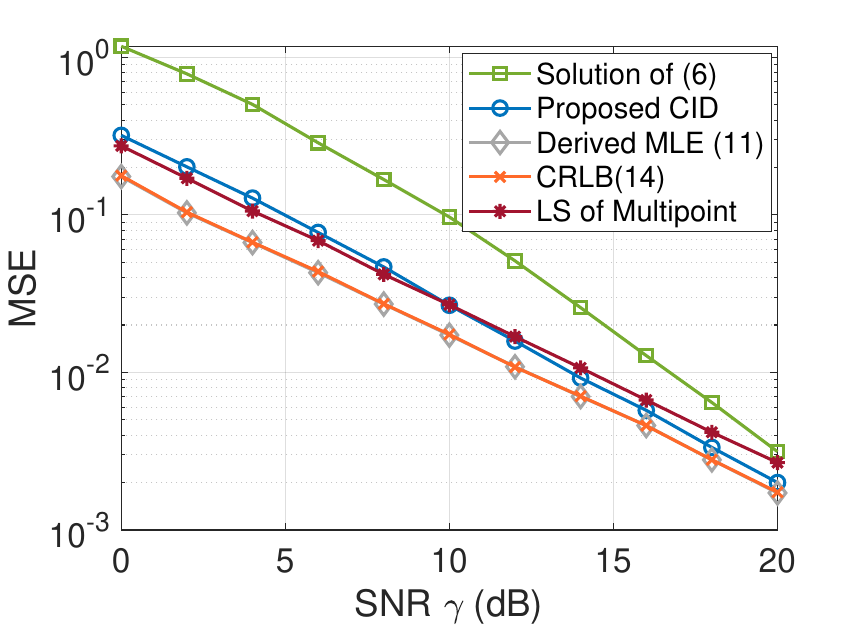}}
        \label{fig:CRLB_d}
	\end{minipage}
 }
 \hspace{-2.0em}
 \subfigure[Angle estimation $\theta$]{
	\begin{minipage}[t]{0.50\linewidth}
	{\includegraphics[width=\textwidth]{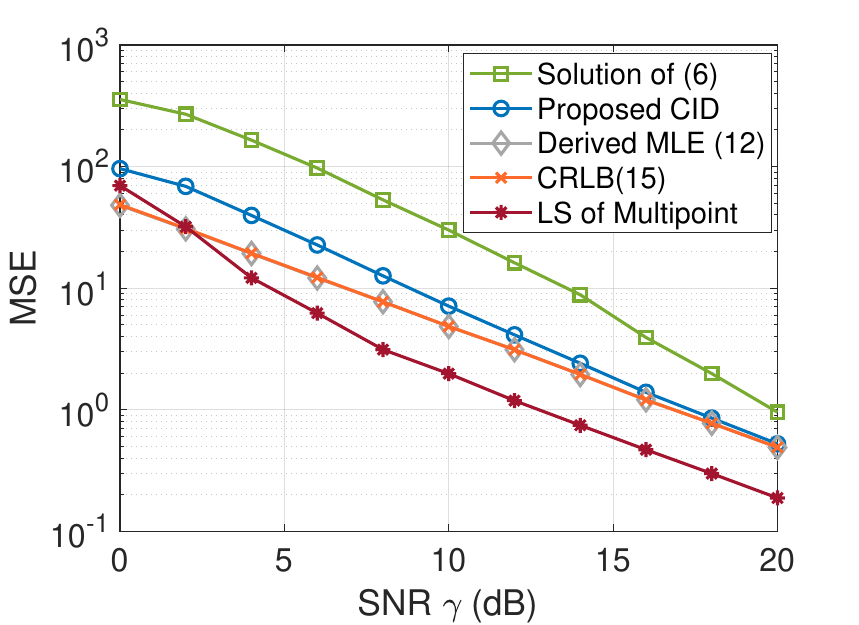}}
        \label{fig:CRLB_theta}
	\end{minipage}
 }
         \caption{Performance comparison between different estimation methods and CRLB  with $N = 8$ measurements, $\Delta\phi_{\mathrm{T}} = 4^{\circ}$.} 
\end{figure}

Figs.~\ref{fig:CRLB_d} and \ref{fig:CRLB_theta} illustrate the MSE performance versus SNR for distance and angle estimation, respectively. The performance of all methods improves with increasing SNR, with the solution of (\ref{eq:solve}) showing the poorest performance. The proposed CID algorithm demonstrates improved accuracy, while the derived MLE achieves near-optimal performance by approaching the theoretical CRLB. 
Compared to multi-point localization methods, our approach demonstrates superior distance estimation while exhibiting lower angular estimation performance. 
This performance difference stems from the inherent characteristics of our pattern diversity approach.
In our system, distance estimation benefits from measurement redundancy as the actual distance remains invariant across different antenna orientations, while each radiation pattern provides an independent signal strength sample. Conversely, angular information is embedded within the relative differences between radiation patterns and obtained through indirect derivation.

\begin{figure}[t]
\subfigure[Distance estimation $d$]{
	\begin{minipage}[t]{0.50\linewidth}
	{\includegraphics[width=1\textwidth]{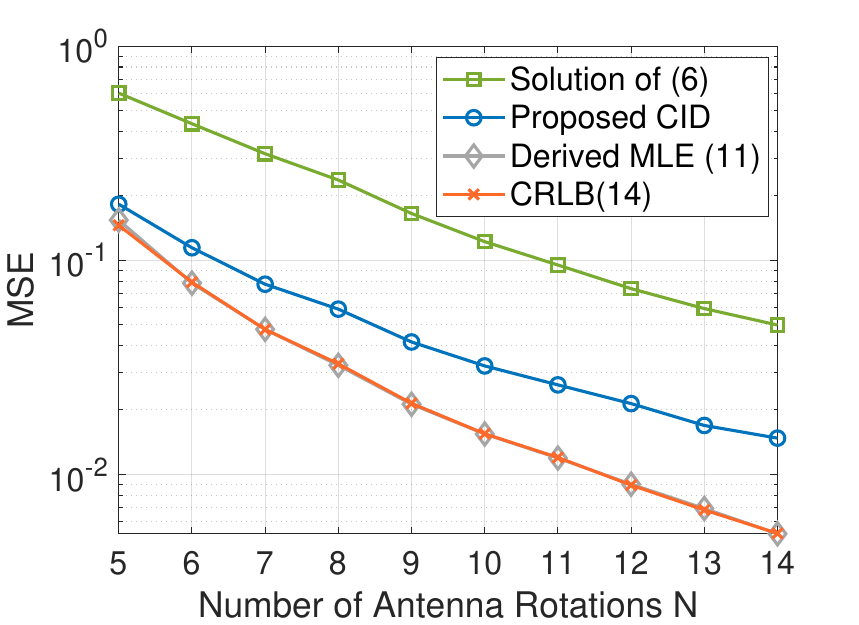}}
        \label{fig:N_d}
	\end{minipage}
 }
 \hspace{-2.0em}
 \subfigure[Distance estimation $\theta$]{
	\begin{minipage}[t]{0.50\linewidth}
	{\includegraphics[width=\textwidth]{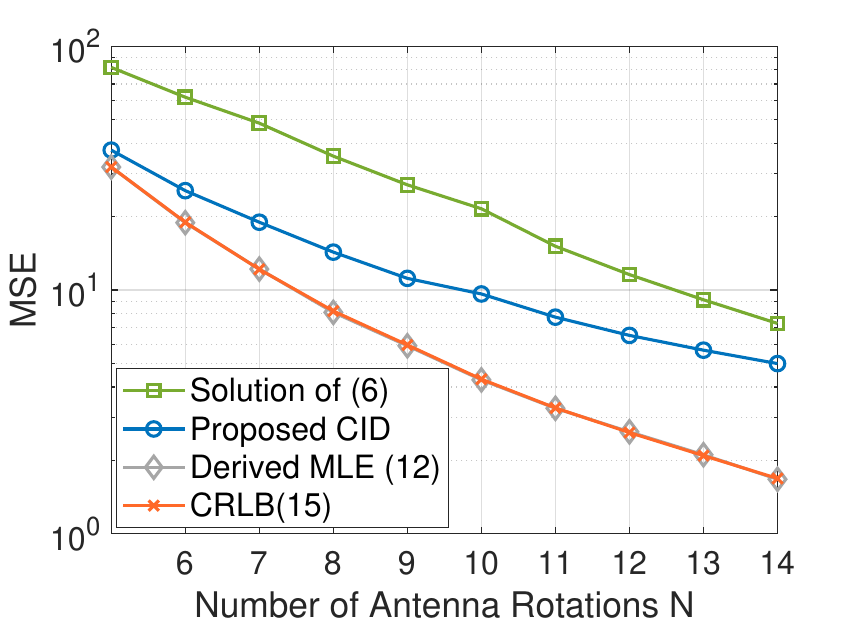}}
        \label{fig:N_theta}
	\end{minipage}
 }
         \caption{Performance comparison between different estimation methods  and CRLB with $\textrm{SNR} = 10$ dB , $\Delta\phi_{\mathrm{T}} = 3^{\circ}$. } 
\end{figure}

The effect of antenna rotation count on estimation accuracy is examined in Figs.~\ref{fig:N_d} and~\ref{fig:N_theta}. Results show enhanced performance for all methods as $N$ increases from 7 to 14, with the MLE maintaining near-CRLB performance throughout. 
These experimental results validate our theoretical analysis \textbf{Propositions \ref{theory:hat_d} and \ref{theory:hat_theta}}, confirming the positive correlation between estimation accuracy and both SNR and measurement count.

\begin{figure}[t] 
\subfigure[Distance estimation $d$]{
	\begin{minipage}[t]{0.50\linewidth}
	{\includegraphics[width=1\textwidth]{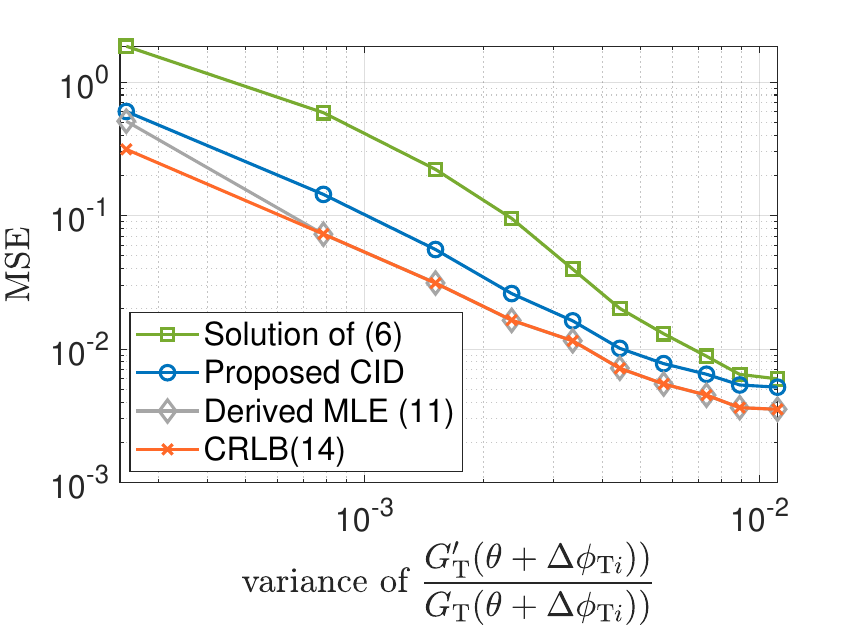}}
        \label{fig:deltaphi_d}
	\end{minipage}
 }
 \hspace{-2.0em}
 \subfigure[Distance estimation $\theta$]{
	\begin{minipage}[t]{0.50\linewidth}
	{\includegraphics[width=\textwidth]{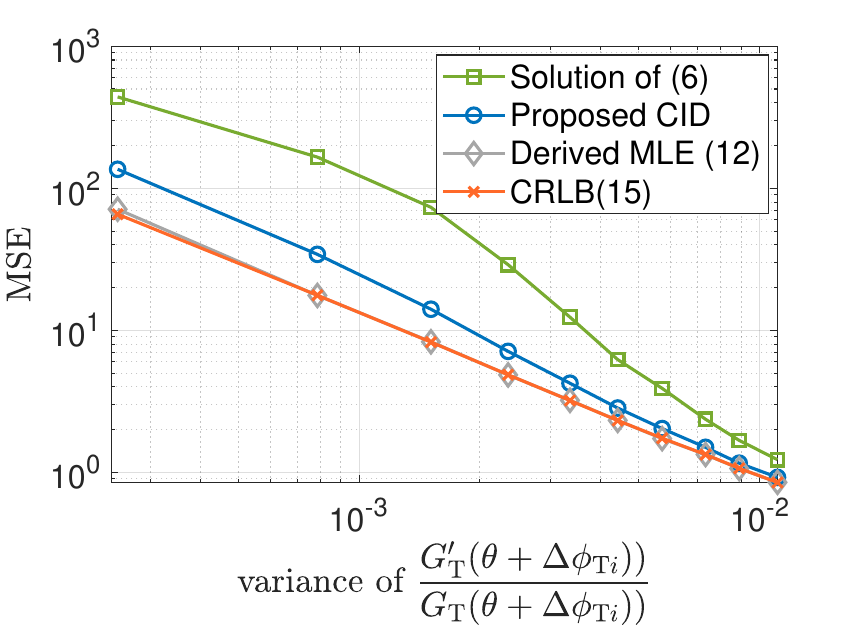}}
        \label{fig:deltaphi_theta}
	\end{minipage}
 }
    \caption{Estimation variance for diatance $\hat{d}$ and angle $\hat{\theta}$ versus variance of $\mathcal{G}$ with SNR = 10 dB, $N=8$ measurements.}
\end{figure}

Figs.\ref{fig:deltaphi_d} and \ref{fig:deltaphi_theta} examine the relationship between distance and angular estimation performance and the variance of $\mathcal{G}$ to quantify the impact of antenna radiation patterns. The results reveal that larger variance in $\mathcal{G}$ leads to improved  estimation accuracy for all methods, which  validates the theoretical analysis in \textbf{Propositions \ref{theory:hat_theta}}.

%% file: body/C5.tex
\section{Conclusion} \label{book:con}
In this paper, we presented a  cost-effective positioning approach leveraging antenna radiation pattern characteristics based on RSSI and proposed corresponding algorithms. Through antenna rotation, we leverages asymmetric radiation patterns to achieve pattern diversity. Based on this system, we make theoretical analysis and simulations, established three fundamental proposition characterizing the estimation accuracy bounds.
Our simulations demonstrates that the derived MLE method achieves near-optimal performance by approaching the CRLB. The proposed CID algorithm demonstrates robust performance across diverse antenna configurations, including both mechanical rotation and electronically reconfigurable patterns. The experimental results validate our theoretical findings that estimation accuracy improves with increased SNR, antenna rotation count, and steeper variations in the antenna radiation pattern.
And compared to multi-point localization methods, our approach demonstrates superior distance estimation while exhibiting lower angular estimation performance. 
To enhance practical applicability, we developed a two-position measurement strategy that eliminates dependence on receiving antenna patterns while maintaining positioning accuracy. This approach provides a promising solution for indoor robot position applications where both accuracy and system simplicity are essential considerations.

%% file: body/appendix.tex
\vspace{-1mm}
\appendices % 多附录

\color{black}
\section{Proof of Proposition \ref{theory:hat_d}} \label{book:hat_d}
Owing to  $\frac{\partial M_i(d, \theta)}{\partial d} = \frac{-10n}{d\ln10}$ is a constant independent of $i$ and not equal to zero,
        (\ref{eq:over_d}) is equivalent to
        \begin{align}
        \small
            \sum_{i=1}^{N}& R(\phi_{\textrm{T}_i})  - \sum_{i=1}^{N}  M_i(\hat{d}, \hat{\theta}) = 0 \label{eq:15} .
        \end{align}
        By substituting  (\ref{eq:Ri}) into (\ref{eq:15}):
        \begin{align}  \label{444}
        \small 
        \frac{\hat{d}^N}{\prod_{i=1}^{N}{\sqrt[n]{A^2G_{\textrm{T}}(\hat{\theta}+\bigtriangleup\phi_{\textrm{T}_i})G_{\textrm{R}}}}}\!=\! 
        \frac{{d_0}^N \cdot \frac{1}{10n} X_\sigma ^{'}}{\prod_{i=1}^{N}{\sqrt[n]{A^2G_{\textrm{T}}({\theta_0}+\bigtriangleup\phi_{\textrm{T}_i})G_{\textrm{R}}}}}. 
        \end{align}
        % From (\ref{444}), the estimated distance $\hat{d}$ is derived as
        % 【公式里面尽量不要we做主语，你可以多用被动语态，The estimated distance is derived as:】
        % 【有公式的部分你都检查一下，我就不一个一个写了，三个proof里面基本都是we做主语】
        \begin{align}
        \small
            &  \hat{d} = d_0 \cdot 10^{\frac{1}{10n\cdot N} X_\sigma ^{'}} \cdot \left(\prod_{i=1}^{N}{\Gamma(\hat{\theta})}\right)^{\frac{1}{n\cdot N}} .
        \end{align}
When $\hat{\theta} = \theta_0$, the expression simplifies to $\hat{d} = d_0 \cdot 10^{\frac{1}{10n\cdot N} X_\sigma ^{'}}$, yielding $\log_{10} \hat{d} = \log_{10} d_0 + \frac{1}{10n \cdot N}X_{{\sigma}}^{'}$.  Consequently, $\log_{10} \hat{d}$ follows a normal distribution: $\log_{10} \hat{d} \sim \mathcal{N}\left(\log_{10} d_0,\frac{\sigma^2}{100n^2 \cdot N} \right)$. Through the probability density transformation theorem, it can be concluded that
\begin{align}
\small
f_{\hat{d}}(h) &= f_{\log_{10} \hat{d}}(\log_{10} h) \cdot \left| \frac{d(\log_{10} h)}{d h} \right|\\
% &\log_{10} \hat{d} = \log_{10} d_0 - \frac{1}{20N}X_{{ \sigma}_i} \notag   \\
&= \frac{10n\cdot \sqrt{N}}{\sigma h(\ln10)\sqrt{2\pi }} { e^{-\frac{50n^2 \cdot N(log_{10}h-log_{10}d_0)^2}{{\sigma^2}} }} .
\end{align}
Then:
\begin{align}
&E[\hat{d}] = \int_{0}^{+\infty} \frac{10n\cdot \sqrt{N}}{\sigma(\ln10)\sqrt{2\pi}} {e^{-\frac{50n^2 \cdot N(log_{10}h-log_{10}d_0)^2}{{\sigma^2}} }}dh \\
&\overset{(a)}{=} \frac{d_0}{\sqrt{2\pi}}\int_{- \infty}^{+\infty} \mathrm{e}^{-\frac{1}{2}t^2 + \frac{\sigma}{10 n} t \cdot \ln10}dt   \\
&=\frac{d_0\cdot e^{\frac{1}{2}(\frac{\sigma}{10n \cdot \sqrt{N}} \ln 10)^2}}{\sqrt{2\pi}}\int_{-\infty}^{+\infty} \mathrm{e}^{-\frac{1}{2}(t - \frac{\sigma}{10n} \ln10)^2  }dt \\
&=d_0\cdot10^{ \frac{\sigma^2}{200n^2 \cdot N} \ln10}. \label{eq:bias} 
\end{align}
where (a) represents: $t = \frac{10n \cdot \sqrt{N}(log_{10}h - log_{10}d_0)}{\sigma}$. 

\color{black}

\section{Proof of Proposition \ref{theory:hat_theta}} \label{book:hat_theta}

Let $\frac{\partial M_i(d, \theta)}{\partial \theta}|_{\theta = \hat{\theta}} = \frac{10}{\ln10}\cdot \frac{G_{\textrm{T}}^{'}(\hat{\theta}+\bigtriangleup\phi_{\textrm{T}_i})}{G_{\textrm{T}}(\hat{\theta}+\bigtriangleup\phi_{\textrm{T}_i})}$ be abbreviated as $K_i$,  $\sum_{i=1}^{N}K_i$ be abbreviated as $\mathcal{H}$ and ${\frac{G_{\textrm{T}}(\hat{\theta}+\bigtriangleup\phi_{\textrm{T}_i})}
{ G_{\textrm{T}}(\theta_0+\bigtriangleup\phi_{\textrm{T}_i})}}$ be abbreviated as $\Gamma_i(\hat{\theta})$. 
Substituting the expression of $R(\phi_{\textrm{T}_i}), M_i(d,\theta)$  into (\ref{eq:over_theta}), we have
     \begin{align}
     \small
        &\sum_{i=1}^{N} 10n\cdot \log_{10}\left(\frac{\hat{d}}{d_0} \cdot \sqrt[n] {\frac{1}{\Gamma_i(\hat{\theta})}}\right)^{K_i}+K_i X_{\sigma i} = 0  \label{eq:222}.
     \end{align}
By applying the logarithmic addition rule, ($\ref{eq:222}$) simplifies to
     \begin{align}
     \tiny
      10n \cdot  \log_{10}\left((\frac{\hat{d}}{d_0})^{\mathcal{H}}\cdot  \prod_{i=1}^{N}  \Gamma_i(\hat{\theta})^{-\frac{K_i}{n}}\right) + K_i X_{{ \sigma}_i}= 0  \label{eq:333}.
     \end{align}
Substituting (\ref{eq:E(d)}) into ($\ref{eq:333}$), we obtain 
     {\small
     \begin{align}
     10n \log_{10}\left(\prod_{i=1}^{N}  \Gamma_i(\hat{\theta})^{\frac{\mathcal{H}-N\cdot K_i}{n\cdot N}}\right)\!+\!\sum_{i=1}^{N} (K_i-\frac{\mathcal{H}}{N} )X_{{\sigma}_i}\!=\!0 .
     \end{align}
     }
$\Gamma_i(\hat{\theta}) \neq 1$ when the estimated angle differs from its true value ($\hat{\theta} \neq \theta_0$). Due to the monotonic nature of the exponential function, the estimation bias magnitude is characterized by $|\frac{\mathcal{H}}{n\cdot N} - \frac{K_i}{n}|$. This term corresponds to the variance of $K_i$, indicating that increased variance of  $\mathcal{G}$ enhances the system's sensitivity to estimation errors.

\color{black}

\section{Proof of Proposition \ref{theory:CRLB}} \label{book:CRLB}
Considering the bias in distance estimation, the calculation of the CRLB for this method requires the use of the following formula from \cite{10.5555/573302}
\begin{align}
\small
&{E}_R\left[(\hat{\alpha}(\mathbf{R}) - \alpha)(\hat{\alpha}(\mathbf{R}) - \alpha)^T\right] \notag \\
&\geq \beta(\alpha)\beta(\alpha)^T 
+ \frac{\partial {E}_R[\hat{\alpha}(\mathbf{R})]}{\partial \alpha^T}
\mathbf{J}_D^{-1}(\alpha)
\frac{\partial {E}_R[\hat{\alpha}(\mathbf{R})]}{\partial \alpha} \label{eq:bias_CRLB},
\end{align}
where $\alpha = [d,\theta]^T$ denotes  the parameter vector to be estimated,  $\hat{\alpha} = [\hat{d},\hat{\theta}]^T$ as the parameter estimate, $\mathbf{R} = [R (\phi_{\textrm{T}_0}),R (\phi_{\textrm{T}_1})...R (\phi_{\textrm{T}_N})] $ as the  random observation vector characterized by a conditional probability density function $p(\mathbf{R}; {\alpha})$  which is equal to $\L(d,\theta)$.
And $\beta(\alpha) = {E}_R\left[\hat{\alpha}(\mathbf{R}) - \alpha\right]$ is defined as the estimation error. 
$\mathbf{J}_D(\alpha)$ is the Fisher Information Matrix (FIM) given by:
\begin{align}
\small
 \mathbf{J}_D({\alpha}) = -{E}_R\left[
{\frac{\partial}{\partial {\alpha}} \ln p(\mathbf{R}; {\alpha}) 
\frac{\partial}{\partial {\alpha}^T} \ln p(\mathbf{R};{\alpha})}
\right] \label{eq:FIM}.  
\end{align}
According to (\ref{eq:FIM}), given the Gaussian observations, elements of the FIM are derived as
\begin{align}
\small
  \left[\mathbf{J}_D\right]_{1,1} &
=  -{E}_R\left[\frac{\partial^2 \ln L(\mathbf{R};d,\theta)}{\partial d^2}  \right] 
% &= {E}_R\left[\frac{1}{\sigma^4} \sum_{i=1}^{N} [R(\phi_{\textrm{T}_i}) - M_i(d, \theta)]^2\cdot({\frac{\partial M_i(d, \theta)}{\partial d}})^2  \right] \notag\\
= \frac{100n^2 \cdot N}{\sigma^2\cdot (d\ln10)^2} ,\\
  \left[\mathbf{J}_D\right]_{1,2} &= \left[\mathbf{J}_D\right]_{2,1} 
=  -{E}_R\left[\frac{\partial^2 \ln L(\mathbf{R};d,\theta)}{\partial d \partial \theta}  \right] \notag\\
% &= {E}_R\left[\frac{1}{\sigma^4} \sum_{i=1}^{N} [R(\phi_{\textrm{T}_i}) - M_i(d, \theta)]^2\cdot{\frac{\partial M_i^2(d, \theta)}{\partial d \partial \theta}}  \right] \notag\\
&=\frac{1}{\sigma^2}\sum_{i=1}^{N} \frac{-10n}{d \cdot\ln 10}\cdot K_i ,\\
\left[\mathbf{J}_D\right]_{2,2} &
= - {E}_R\left[\frac{\partial^2 \ln L(\mathbf{R};d,\theta)}{\partial \theta^2}  \right] \notag\\
% &= {E}_R\left[\frac{1}{\sigma^4} \sum_{i=1}^{N} [R(\phi_{\textrm{T}_i}) - M_i(d, \theta)]^2\cdot({\frac{\partial M_i(d, \theta)}{\partial \theta}})^2  \right] \notag\\
&=\frac{1}{\sigma^2} \sum_{i=1}^{N} K_i^2.
\end{align}
By applying matrix inverse, we obtain
\begin{align}
\small
& \left[\mathbf{J}_D^{-1}\right]_{1,1} = \frac{\left[\mathbf{J}_D\right]_{2,2}}
{\left[\mathbf{J}_D\right]_{1,1} \left[\mathbf{J}_D\right]_{2,2} - \left[\mathbf{J}_D\right]_{1,2}^2} \notag \\
&=\frac{\frac{\sigma^2 d^2 \ln^2 10}{100n^2} \cdot \sum_{i=1}^{N} K_i^2}{N \cdot \sum_{i=1}^{N} K_i^2 -
\left( \sum_{i=1}^{N} K_i\right)^2} ,\\
&\left[\mathbf{J}_D^{-1}\right]_{2,2} = \frac{\left[\mathbf{J}_D\right]_{1,1}}
{\left[\mathbf{J}_D\right]_{1,1} \left[\mathbf{J}_D\right]_{2,2} - \left[\mathbf{J}_D\right]_{1,2}^2} \notag\\
&=\frac{ N }{N\cdot \sum_{i=1}^{N} K_i^2 -\left( \sum_{i=1}^{N} K_i \right)^2} ,\\
&\left[\mathbf{J}_D^{-1}\right]_{1,2} = \left[\mathbf{J}_D^{-1}\right]_{2,1}=\frac{- \left[\mathbf{J}_D\right]_{1,2}}
{\left[\mathbf{J}_D\right]_{1,1} \left[\mathbf{J}_D\right]_{2,2} - \left[\mathbf{J}_D\right]_{1,2}^2} \notag\\
&=\frac{\frac{\sigma^2 d \ln 10}{10n} \cdot \sum_{i=1}^{N} K_i^2}{N \cdot \sum_{i=1}^{N} K_i^2 -
\left( \sum_{i=1}^{N} K_i\right)^2} .
\end{align}
Then leveraging the distance estimation formulation (\ref{eq:E(d)}), we have:
\begin{align}
\beta_d &= E\left(d \cdot 10^{\frac{1}{10n \cdot N} X_\sigma ^{'}} \cdot (\prod_{i=1}^{N}{{\Gamma(\hat{\theta})}})^{\frac{1}{n\cdot N}} - d \right) \label{eq:beta_d},\\
\frac{\partial \beta_d}{\partial d} &= E\left( 10^{\frac{1}{10n \cdot N} X_\sigma ^{'}} \cdot (\prod_{i=1}^{N}{{\Gamma(\hat{\theta})}})^{\frac{1}{n\cdot N}} - 1\right)\label{eq:pbeta_pd}, \\
 \frac{\partial\beta_d}{\partial \theta} &=d \cdot E\left( 10^{\frac{1}{10n \cdot N} X_\sigma ^{'}} \cdot \frac{\partial (\prod_{i=1}^{N}{{\Gamma(\hat{\theta})}})^{\frac{1}{n\cdot N}}}{\partial \theta}\right).
\end{align}